# Modeling a foreign exchange rate using moving average of Yen-Dollar market data

Takayuki Mizuno[1], Misako Takayasu[1], Hideki Takayasu[2]

[1] Department of Computational Intelligence and Systems Science, Interdisciplinary Graduate School of Science and Engineering, Tokyo Institute of Technology, 4259 Nagatsuta-cho, Midori-ku, Yokohama 226-8502, Japan
[2] Sony Computer Science Laboratories, 3-14-13 Higashi-Gotanda, Shinagawa-ku, Tokyo 141-0022, Japan

**Summary.** We introduce an autoregressive-type model with self-modulation effects for a foreign exchange rate by separating the foreign exchange rate into a moving average rate and an uncorrelated noise. From this model we indicate that traders are mainly using strategies with weighted feedbacks of the past rates in the exchange market. These feedbacks are responsible for a power law distribution and characteristic autocorrelations of rate changes.

## 1. Introduction

The probability densities of rate changes of foreign exchange markets generally have fat tails compared with the normal distribution and the volatility always shows a long autocorrelation [1]. In order to clarify the mechanism of these nontrivial behaviors, we introduce an auto-regressive type model with self-modulation effects for the exchange rate by using the new technique of separating moving average rates and residual uncorrelated noises [2,3]. We are going to show that these nontrivial behaviors are caused by traders' strategies with weighted feedbacks of the past rates. In this paper we use a set of tick-by-tick data provided by CQG for the yen-dollar exchange rates from 1989 to 2002.

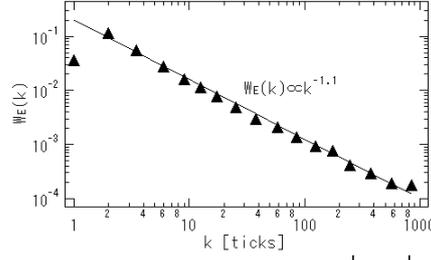

Fig.1 Weight factors $w_\varepsilon(k)$ of the absolute value $|\varepsilon(t)|$ of the yen-dollar rate. The line indicates a power function $w_\varepsilon(k) \propto k^{-1.1}$.

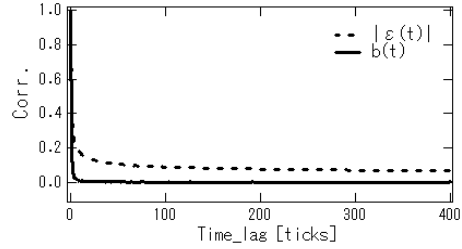

Fig.2 Autocorrelations of the absolute value $|\varepsilon(t)|$ and the factor $b(t)$.

## 2. The best moving average

Traders are generally predicting future exchange rates using various types of moving averages. We first introduce so-called the best moving average rate that separates uncorrelated noises from the market data.

A foreign exchange rate $P(t+1)$ is generally separable into a moving average rate $\overline{P}(t)$ and its residue $\varepsilon(t)$,

$$P(t+1) = \overline{P}(t) + \varepsilon(t), \quad (1)$$
$$\overline{P}(t) = \sum_{k=1}^{K} w_P(k) \cdot P(t-k+1), \quad (2)$$

where $w_P(k)$ gives the weight factors where the time is measured by ticks. By tuning the weight factors we tried to find the best set of weights that makes the autocorrelation of the term $\varepsilon(t)$ almost zero. It is found that such condition is satisfied generally by weights which decay nearly exponentially with a characteristic time about a few minutes.

Although the correlation of $\varepsilon(t)$ is nearly zero, its absolute value shows a long autocorrelation [2]. In order to characterize this stochastic dynamics we also separate the absolute value $|\varepsilon(t+1)|$ into a moving average $\langle|\varepsilon(t)|\rangle$ and an uncorrelated noise term, $b(t)$. We apply an autoregressive process to $\log|\varepsilon(t+1)|$ as follows,

$$\log|\varepsilon(t+1)| = \log\langle|\varepsilon(t)|\rangle + \log b(t), \quad (3)$$

$$\log\langle|\varepsilon(t)|\rangle = \sum_{k=1}^{K'} w_\varepsilon(k) \cdot \log|\varepsilon(t-k+1)|, \quad (4)$$

where $w_\varepsilon(k)$ is the weight factor which is estimated from the foreign exchange data. The weight factors $w_\varepsilon(k)$ of the yen-dollar rate decay according to power law $w_\varepsilon(k) \propto k^{-1.1}$ with a characteristic time about a few minutes as shown in Fig.1. The autocorrelation of the term $b(t)$ becomes nearly zero as shown in Fig.2. Namely, the fluctuation of the logarithm of absolute value of $\varepsilon(t)$ can be approximated by an autoregressive type stochastic process.

From these results, we find that the characteristic time of the best moving average is generally about a minute, namely, most traders are expected to be watching only very latest market data of order of a few minutes.

## 3. Self modulation process for foreign exchange rate

As a mathematical model of foreign exchange market that is directly compatible with the tick-by-tick data, we now introduce an auto-regressive type model with self-modulation effects as follows,

$$\begin{cases} P(t+1) = \overline{P}(t) + \varepsilon(t) & (5) \\ \varepsilon(t+1) = \alpha(t) \cdot b(t) \cdot \langle|\varepsilon(t)|\rangle + f(t) & (6) \end{cases},$$

where the moving averages $\overline{P}(t)$ and $\langle|\varepsilon(t)|\rangle$ are given by Eqs.(2) and (4), $\alpha(t)$ is chosen randomly from 1 or -1 with probability 0.5. We introduce an additive term $f(t)$ independent of $\langle|\varepsilon(t)|\rangle$ in order to take into account effects such as sudden big news or interventions by the central banks or other uncertain events.

We simulate the rate changes numerically by using Eqs.(5) and (6). In the simulation the noise $b(t)$ is chosen randomly from the observed probability density for "$b$" in Eq.(3). As for the weight function in the moving average in Eq.(5), we apply an exponential function, $w_P(k) = 0.43 e^{-0.35k}$. The external noise factor $f(t)$ is given by a Gaussian noise with the average value 0 and its standard deviation 0.001.

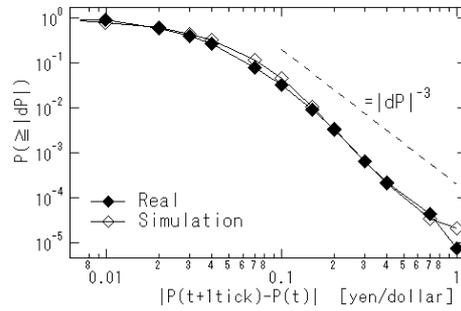

Fig.3 Cumulative distributions of rate changes.

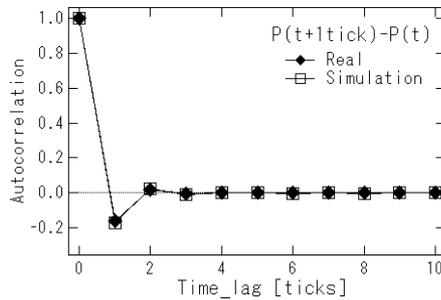 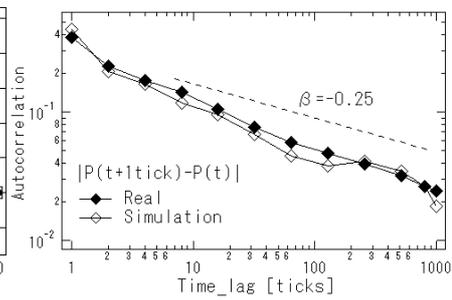

Fig.4 Autocorrelations of rate change.　　Fig.5 Autocorrelations of volatility.

We compare the simulated rates to the real yen-dollar rates. In Fig.3 the cumulative distribution of rate changes $|P(t+1\text{tick}) - P(t)|$ by our simulation is plotted together with the real data. The two graphs fit quite nicely both showing power law behaviors as indicated in the figure.

This power law property can be understood theoretically from the view point of self-modulation process that is a stochastic process of which basic parameters such as the mean value are modulated by the moving average of its own traces [4,5,6]. According to the results of self-modulation processes it is a natural consequence that the resulting market rates show power law properties when the multiplicative factor $b(t)$ in Eq.(6) fluctuates randomly.

The autocorrelation of rate changes and that of the volatility are plotted in Fig.4 and Fig.5, respectively. In both cases the simulation results fit with the real data quite nicely. It should be noted that the functional form of the autocorrelation functions depend on the weight factors $w_P(k)$ and $w_\varepsilon(k)$, and the interesting point is that the weight factors work quite well, namely, the principle of making the residue terms independent is effective.

## 4. Discussion

We introduced a new type of foreign exchange rate equation that describes very short time characteristics of markets consistent with the real data. It is well-known that traders are generally using moving average methods for predicting the future rates. Our model represents this general property of traders by introducing the best weight factors of the moving averages $w_P(k)$, $w_\varepsilon(k)$ and the noise factors $b(t)$ that expresses responses of dealers to the past market rate changes. From our model it is confirmed that this feedback of information is responsible for the power law distribution of rate changes and characteristic autocorrelations of rate changes and volatility.

## Acknowledgement

The authors would like to show appreciation to Hiroyuki Moriya of Oxford Financial Education for providing us the data of high-frequency exchange rate, Prof. Tohru Nakano of Chuo Univ. for stimulating discussions. T. Mizuno is supported by Research Assistant Fellowship of Chuo University, and the Ministry of Education, Science, Sports and Culture, Grant-in-Aid for JSPS Fellows.